\documentclass[pra,twocolumn,showpacs,preprintnumbers,superscriptaddress]{revtex4}
\usepackage{times}
\usepackage{bm}
\usepackage{graphicx}
\usepackage{amsbsy}
\usepackage{amsmath}
\usepackage{amsfonts}
\usepackage{amsthm}
\usepackage{xcolor}

\begin{document}

\theoremstyle{plain}
\newtheorem{theorem}{Theorem}
\newtheorem{lemma}[theorem]{Lemma}
\newtheorem{corollary}[theorem]{Corollary}
\newtheorem{proposition}[theorem]{Proposition}\newtheorem{conjecture}[theorem]{Conjecture}
\theoremstyle{definition}
\newtheorem{definition}[theorem]{Definition}
\theoremstyle{remark}
\newtheorem*{remark}{Remark}
\newtheorem{example}{Example}
\title{On a matrix equality involving partial transposition and its relation to the separability problem}
\author{Vaibhav Soni, Rishone Deshwal, Aayush Garg, Rohit Kumar, Satyabrata Adhikari}
\email{vaibhav.soni2199@gmail.com,rishoneawesome@gmail.com,aayushgarg048@gmail.com,rohitkumar@dtu.ac.in,satyabrata@dtu.ac.in} \affiliation{Delhi Technological University, Shahbad Daulatpur, Main Bawana Road, Delhi-110042,
India}

\begin{abstract}
In matrix theory, a well established relation $(AB)^{T}=B^{T}A^{T}$ holds for any two matrices $A$ and $B$ for which the product $AB$ is defined. Here $T$ denote the usual transposition. In this work, we explore the possibility of deriving the matrix equality $(AB)^{\Gamma}=A^{\Gamma}B^{\Gamma}$ for any $4 \times 4$ matrices $A$ and $B$, where $\Gamma$ denote the partial transposition. We found that, in general, $(AB)^{\Gamma}\neq A^{\Gamma}B^{\Gamma}$ holds for $4 \times 4$ matrices $A$ and $B$ but there exist particular set of $4 \times 4$ matrices for which $(AB)^{\Gamma}= A^{\Gamma}B^{\Gamma}$ holds. We have exploited this matrix equality to investigate the separability problem.  Since it is possible to decompose the density matrices $\rho$ into two positive semi-definite matrices $A$ and $B$ so we are able to derive the separability condition for $\rho$ when $\rho^{\Gamma}=(AB)^{\Gamma}=A^{\Gamma}B^{\Gamma}$ holds. Due to the non-uniqueness property of the decomposition of the density matrix into two positive semi-definte matrices $A$ and $B$, there is a possibility to generalise the matrix equality  for density matrices lives in higher dimension. These results may help in studying the separability problem for higher dimensional and multipartite system. 
\end{abstract}
\pacs{03.67.Hk, 03.67.-a} \maketitle
Keywords: Partial Transpose, Positive semi-definite matrix, Quantum Entanglement
\section{Introduction}
Quantum Entanglement \cite{einstein} is a physical phenomenon that has no classical analogues and thus distinguishes the quantum world from its classical counterparts. Entangled states can be used as a resource which proved to be an efficient quantum channel in comparison to classical resources. The existence of such long range correlations in quantum systems provide optimal success not only in quantum communication \cite{bennett3, ekert,bennett5,bennett1} but also in quantum computation \cite{ekert1}. Since quantum correlation in an entangled state outperform classical resources so it is essential to know whether the prepared quantum state is an entangled state or separable state? Thus one of the important topic in quantum information theory is the detection of entanglement.\\
There are many methods can be found in the literature for the detection of entanglement \cite{horodecki4,guhne2}. The first method to detect entanglement is the partial transposition method and it is introduced by Peres \cite{peres}. If we consider a $2 \otimes 2$ quantum system described by the density operator $\rho$ then the quantum state $\rho$ can be represented as a block matrix
\begin{eqnarray} \rho=
\begin{pmatrix}
  A & B \\
  B^{\dagger} & C
\end{pmatrix}
  \label{block}
\end{eqnarray}
where $A$, $B$, $C$ denote the $2 \otimes 2$ matrices. Then the partial transposition $\Gamma$ of the density matrix is designated as $\rho^{\Gamma}$ and it s defined by
\begin{eqnarray} \rho^{\Gamma}=
\begin{pmatrix}
  A^{T} & B^{T} \\
  (B^{\dagger})^{T} & C^{T}
\end{pmatrix}
  \label{blockPT}
\end{eqnarray}
where $T$ denote the ordinary transposition.\\
Later Horodecki's \cite{horodecki2} proved that the partial transposition method is necessary and sufficient for $2 \otimes 2$ and $2 \otimes 3$ bipartite quantum system. After these seminal works, lot of work had already been done in the context of the detection of two-qubit entanglement. But in spite of these, still there are few questions that remain to be answered. For instance, it is a well known fact that for any two matrices $A$ and $B$,
\begin{eqnarray}
(AB)^{T}=B^{T}A^{T}
\label{trans1}
\end{eqnarray}
holds.\\
The question is whether the above equality (\ref{trans1}) holds for partial transposition also. That is, if $\Gamma$ denote partial transposition then the matrix equality
\begin{eqnarray}
(AB)^{\Gamma}=A^{\Gamma}B^{\Gamma}
\label{partrans1}
\end{eqnarray}
holds?\\
In this work, we would explore the validity of the matrix equality given by (\ref{partrans1}) and then will apply the obtained result in investigating the separability problem for two-qubit system. In particular, we will investigate the matrix equality to study the separability problem of two-qubit X states. Two-qubit X states looks like a alphabet X and it is described by a density operator with seven real-valued parameters. Two-qubit maximally entangled Bell states \cite{nielsen} and the Werner states \cite{werner}  belong to the class of X states. If we consider the higher order matrices and further if we take the entries of the matrices as variable (not constant) then it will be difficult to obtain the analytical expression of the eigenvalue even with software also. In this work, we have taken into account this problem. To investigate this problem, we decompose the matrix into two matrices whose eigenvalues may be obtained in a relatively easier way. This problem is vital in the context of quantum information theory in which, the detection of entangled states is considered as one of the important problem. The problem of detection of entanglement is largely based on finding the eigenvalues. This motivate us to consider this problem.
\section{Is partial transposition of product of two positive semi-definite (PSD) matrices equal to product
of partial transposition of individual PSD matrices?}
In this section, we will investigate the truthfulness of the matrix inequality (\ref{partrans1}).  To investigate, we choose the matrices $P_{1}$ and $P_{2}$ in such a way so that if the relation $(P_{1}P_{2})^{\Gamma}=P_{1}^{\Gamma}P_{2}^{\Gamma}$ holds then it should have some application
in quantum information theory.\\
Let us choose two positive semi-definite matrices $P_{1}$, $P_{2}$ that can be expressed in the form of $2 \times 2$ block matrices as
\begin{eqnarray}
P_{1}=
\begin{pmatrix}
  T_1 & 0 \\
  0 & T_2
  \end{pmatrix} \in \mathbb{C}^{4\times4},
P_{2}=
\begin{pmatrix}
  I & V\\
  V^{\dagger} & I
  \end{pmatrix} \in \mathbb{C}^{4\times4}
  \label{p1p2}
\end{eqnarray}
where $T_1, T_2, V \in \mathbb{C}^{2\times2}$.\\
We choose the block matrices {$T_1, T_2, V$} in such a way that the matrix $P_{1}P_{2}$ will look like
the alphabet X. Therefore, we can always choose the matrices {$T_1, T_2, V$} as
{
\begin{eqnarray} &&T_1=
\begin{pmatrix}
  t_{1} & 0 \\
  0 & t_{2}
  \end{pmatrix},
T_2=
\begin{pmatrix}
  t_{3} & 0 \\
  0 & t_{4}
  \end{pmatrix},
V=
\begin{pmatrix}
  0 & v_{1} \\
  v_{2} & 0
  \end{pmatrix}
  \label{parts1}
\end{eqnarray}}\\
Here we assume that the matrix elements {$t_{1}$, $t_{2}$, $t_{3}$, $t_{4}$} are non-negative real numbers and {$v_{1}$, $v_{2}$} are complex entities.\\
Therefore, the matrices $P_{1}$ and $P_{2}$ can take the form
{
\begin{eqnarray} P_{1}=
  \begin{pmatrix}
  t_{1} & 0 & 0 & 0   \\
  0 & t_{2} & 0 & 0 \\
  0 & 0 & t_3 & 0 \\
  0 & 0 & 0 & t_{4}
  \end{pmatrix},
  P_{2}=
  \begin{pmatrix}
  1 & 0 & 0 & v_1\\
  0 & 1 & v_{2} & 0 \\
  0 & v_2^* & 1 & 0 \\
  v_1^* & 0 & 0 & 1
  \end{pmatrix}
  \label{P1}
\end{eqnarray}}
The eigenvalues of $P_{1}$ and $P_{2}$ are given by {$\{t_{1},t_{2},t_3,t_{4}\}$} and {$\{1+|v_1|,1+|v_{2}|,1-|v_1|,1-|v_{2}|\}$} respectively.
Since {$t_{1}$, $t_{2}$, $t_3$ and $t_{4}$ are assumed to be non-negative real numbers} so the matrices $P_{1}$ and $P_{2}$ represent positive semi-definite matrices only when $|v_1|, |v_2| \leq 1$.\\
The product of the two matrices $P_{1}$ and $P_{2}$ is given by
{
\begin{eqnarray}P_{1}P_{2}=
 \begin{pmatrix}
  t_{1} & 0 & 0 & t_{1}v_{1}\\
  0 & t_{2} & t_{2}v_{2} & 0\\
  0 & t_{3}v_{2}^* & t_{3} & 0\\
  t_{4}v_{1}^* & 0 & 0 & t_{4}
  \end{pmatrix}
  \label{P1P2}
\end{eqnarray}}
It can be easily seen that the matrix $P_{1}P_{2}$ looks like the alphabet X
and hence $P_{1}P_{2}$ is a X-shaped matrix.\\
\textbf{Lemma-1:} The matrix $P_{1}P_{2}$ will be hermitian and PSD if $t_1 = t_4$, and  $t_2=t_3$.\\
\textbf{Proof:} If $t_1 = t_4$, and  $t_2=t_3$ then the matrix $P_{1}P_{2}$ given in (\ref{P1P2}) will then reduces to
{
\begin{eqnarray}P_{1}P_{2}=
 \begin{pmatrix}
  t_{1} & 0 & 0 & t_{1}v_{1}\\
  0 & t_{2} & t_{2}v_{2} & 0\\
  0 & t_{2}v_{2}^* & t_{2} & 0\\
  t_{1}v_{1}^* & 0 & 0 & t_{1}
  \end{pmatrix}
  \label{newP1P2}
\end{eqnarray}}
\\
Since it can be seen that $P_{1}P_{2}=(P_{1}P_{2})^{\dagger}$ so the matrix $P_{1}P_{2}$ is hermitian. Also, the eigenvalues of the matrix $P_{1}P_{2}$ are given by
\begin{eqnarray}
 t_1(1-|v_1|),\ \  t_2(1-|v_2|)
\label{trans}
\end{eqnarray}
Since all eigenvalues are non-negative so the matrix $P_{1}P_{2}$ represent
a PSD.\\
Next, we consider the partial transposition of the matrix $P_{1}P_{2}$, which is given by
{
\begin{eqnarray} (P_{1}P_{2})^{\Gamma}=
  \begin{pmatrix}
  t_1 & 0 & 0 & t_2v_2\\
  0 & t_2 & t_1v_1 & 0\\
  0 & t_1v_1^* & t_2 & 0\\
  t_2v_2^* & 0 & 0 & t_1
  \end{pmatrix}
  \label{pt1}
\end{eqnarray}}
The minimum eigenvalue of the matrix $(P_{1}P_{2})^{\Gamma}$ is given by
\begin{eqnarray}
\lambda_{min}((P_{1}P_{2})^{\Gamma})=min\{\lambda^{+},\lambda^{-},\mu^{+},\mu^{-}\}
\label{trans}
\end{eqnarray}
where\\
{
$\lambda^{+}= t_1 + t_2|v_2|,\\
\lambda^{-}= t_1 - t_2|v_2|,\\
\mu^{+}= t_2 + t_1|v_1|,\\
\mu^{-}= t_2 - t_1|v_1|.\\$}
The partial transposition of the matrices $P_{1}$ and $P_{2}$ are given by
{
\begin{eqnarray} P_{1}^{\Gamma}=
  \begin{pmatrix}
  t_{1} & 0 & 0 & 0 \\
  0 & t_{2} & 0 & 0 \\
  0 & 0 & t_2 & 0 \\
  0 & 0 & 0 & t_{1}
  \end{pmatrix},
  P_{2}^{\Gamma}=
  \begin{pmatrix}
  1 & 0 & 0 & v_{2} \\
  0 & 1 & v_1 & 0 \\
  0 & v_1^* & 1 & 0 \\
  v_2^* & 0 & 0 & 1
  \end{pmatrix}
  \label{P1}
\end{eqnarray}}
Now we are in a position to discuss the nature of the matrix expression $(P_{1}P_{2})^{\Gamma}-P_{1}^{\Gamma}P_{2}^{\Gamma}$. In order to
investigate, let us calculate the matrix $P_{1}^{\Gamma}P_{2}^{\Gamma}$, which is given by
{
\begin{eqnarray}P_{1}^{\Gamma}P_{2}^{\Gamma}=
 \begin{pmatrix}
  t_1 & 0 & 0 & t_1v_2\\
  0 & t_2 & t_2v_1 & 0\\
  0 & t_2v_1^* & t_2 & 0\\
  t_1v_2^* & 0 & 0 & t_1
  \end{pmatrix}
  \label{pt2}
\end{eqnarray}}
Therefore, the matrix $(P_{1}P_{2})^{\Gamma}-P_{1}^{\Gamma}P_{2}^{\Gamma}$ is given by
{
\begin{eqnarray}(P_{1}P_{2})^{\Gamma}-P_{1}^{\Gamma}P_{2}^{\Gamma}=
 \begin{pmatrix}
  0 & 0 & 0 & rv_2 \\
  0 & 0 & -rv_1 & 0 \\
  0 & -rv_1^* & 0 & 0 \\
  rv_2^* & 0 & 0 & 0
  \end{pmatrix}
  \label{matrix1}
\end{eqnarray}}
where {$r=(t_{2}-t_{1})$}\\
The eigenvalues of the matrix (\ref{matrix1}) are given by\\
{
$e_{1}=r\ |v_1|$, $e_{2}=-r\ |v_1|$,\\
$e_{3}=r\ |v_2|$, $e_{4}=-r\ |v_2|$.\\}
In general, the matrix $(P_{1}P_{2})^{\Gamma}-P_{1}^{\Gamma}P_{2}^{\Gamma}$ is indefinite.\\
Therefore, we can state the following theorem:\\
\textbf{Theorem-1:} In general, if $X \in \mathcal{C}^{4\times4}$ and $Y \in
\mathcal{C}^{4\times4}$ be any two positive semi-definite matrices then
\begin{eqnarray}
(XY)^{\Gamma}\neq X^{\Gamma}Y^{\Gamma}
 \label{result1}
\end{eqnarray}
$\Gamma$ denotes the partial transposition.\\
It is now important to discuss a particular case for which $(XY)^{\Gamma}= X^{\Gamma}Y^{\Gamma}$ holds.
\subsection{Special Case}
Recalling the X-matrix which is described by $P_{1}P_{2}$ and review the expression $(P_{1}P_{2})^{\Gamma}-P_{1}^{\Gamma}P_{2}^{\Gamma}$ given by (\ref{matrix1}). We observe that it is possible to impose some condition on $t_{1}$, $t_{2}$ so that $(P_{1}P_{2})^{\Gamma}-P_{1}^{\Gamma}P_{2}^{\Gamma}$ can be a null matrix and thus we can achieve the matrix equality relation $(P_{1}P_{2})^{\Gamma}=P_{1}^{\Gamma}P_{2}^{\Gamma}$ under the imposed condition.\\
{If we choose $t_{1}=t_{2}$, then $r=0$ holds and we have}
\begin{eqnarray}
(P_{1}P_{2})^{\Gamma}-P_{1}^{\Gamma}P_{2}^{\Gamma}= 0
\label{zeromat}
\end{eqnarray}
where $0$ represent a $4 \times 4$ null matrix.\\
Thus, we have the matrix equality relation
{\begin{eqnarray}
\delta=(P_{1}P_{2})^{\Gamma}=P_{1}^{\Gamma}P_{2}^{\Gamma}=
 \begin{pmatrix}
  t_{1} & 0 & 0 & t_{1}v_{2}\\
  0 & t_{1} & t_{1}v_{1} & 0 \\
  0 & t_{1}v_{1}^* & t_{1} & 0 \\
  t_{1}v_{2}^* & 0 & 0 & t_{1}
  \end{pmatrix}
 \label{equalcond}
\end{eqnarray}}
\section{The matrix equality $(P_{1}P_{2})^{\Gamma}=P_{1}^{\Gamma}P_{2}^{\Gamma}$ and the separability problem}
In this section, we derive the condition of separability assuming the validity of the matrix equality $(P_{1}P_{2})^{\Gamma}=P_{1}^{\Gamma}P_{2}^{\Gamma}$, where $\Gamma$ denote the partial transposition.\\
It has been shown that any positive semi-definite matrix can be decomposed as the product of two positive semi-definite matrices
\cite{cui}. Therefore, a quantum state described by the density operator $\rho$ can be decomposed as
\begin{eqnarray}
\rho= P_{1}P_{2} \label{decompositionrho1}
 \end{eqnarray}
If we assume that the matrix equality $(P_{1}P_{2})^{\Gamma}=P_{1}^{\Gamma}P_{2}^{\Gamma}$ holds then we have
\begin{eqnarray}
\rho^{\Gamma}= (P_{1}P_{2})^{\Gamma}= P_{1}^{\Gamma}P_{2}^{\Gamma}\label{decompositionrho2}
 \end{eqnarray}
\textbf{Theorem-2:} If $P_{1}$ and $P_{2}$ be two positive definite matrices such that $\rho^{\Gamma}= (P_{1}P_{2})^{\Gamma}=P_{1}^{\Gamma}P_{2}^{\Gamma}$, where $\rho^{\Gamma}$ denote the partial transposition of the quantum state in $2\otimes 2$ system described by the density matrix $\rho$. Further, if it satisfies the inequality
\begin{eqnarray}
&&81(det(P_{1}^{\Gamma}))^{3}(det(P_{2}^{\Gamma}))^{3}+512(det(P_{1}^{\Gamma}))(det(P_{2}^{\Gamma}))\nonumber\\ \leq && \frac{2^{11}}{9}
\label{sepcond1}
\end{eqnarray}
Then the state $\rho$ is separable.\\
\textbf{Proof:} Let us start with the result by Pablo Tarazaga \cite{tarazaga}, which states that
if $Tr(A)>(n-1)^{\frac{1}{2}}\|A\|_{2}$, then $A$ is a full rank matrix of order n and also it is positive definite. $\|A\|_{2}$ denote the Frobenius norm.\\
Let us consider a $2\otimes 2$ quantum state described by the density operator $\rho$. We then apply the result \cite{tarazaga} on  $\rho^{\Gamma}$. Therefore, if
\begin{eqnarray}
Tr(\rho^{\Gamma})>3^{\frac{1}{2}}\|\rho^{\Gamma}\|_{2}
\label{sepcond2}
\end{eqnarray}
holds then $\rho^{\Gamma}$ is full rank and also positive definite. Hence the state $\rho$ is separable.\\
Further since $Tr(\rho^{\Gamma})=1$ and using (\ref{decompositionrho2}), the inequality (\ref{sepcond2}) reduces to
\begin{eqnarray}
1&>&3^{\frac{1}{2}}\|P_{1}^{\Gamma}P_{2}^{\Gamma}\|_{2}\nonumber\\&\geq&
3^{\frac{1}{2}}\sigma_{max}(P_{1}^{\Gamma}P_{2}^{\Gamma})
\nonumber\\&\geq& \frac{9}{3^{\frac{3}{2}}}det(P_{1}^{\Gamma}P_{2}^{\Gamma})
[1+\frac{81}{512}(det(P_{1}^{\Gamma}P_{2}^{\Gamma}))^{2}]
\label{sepcond3}
\end{eqnarray}
where $\sigma_{max}(P_{1}^{\Gamma}P_{2}^{\Gamma})$ denote the maximum singularvalue of $P_{1}^{\Gamma}P_{2}^{\Gamma}$. The second last and final inequality follows from \cite{zhan,sheng}.\\
After a little bit simplification, (\ref{sepcond3}) reduces to
\begin{eqnarray}
&&81(det(P_{1}^{\Gamma}))^{3}(det(P_{2}^{\Gamma}))^{3}+512(det(P_{1}^{\Gamma}))(det(P_{2}^{\Gamma}))\nonumber\\ \leq && \frac{2^{11}}{9}
\label{sepcond4}
\end{eqnarray}
Hence proved.\\
\textbf{Corollary-1:} If $P_{1}$ and $P_{2}$ be two positive definite matrices of order $mn (m\geq 3,n\geq 3)$ such that $\varrho^{\Gamma}= (P_{1}P_{2})^{\Gamma}=P_{1}^{\Gamma}P_{2}^{\Gamma}$, where $\rho^{\Gamma}$ denote the partial transposition of the quantum state $\varrho$ in $m\otimes n$ system. Further, if it satisfies the inequality
\begin{eqnarray}
&&(n-1)^{n}(det(P_{1}^{\Gamma}))^{3}(det(P_{2}^{\Gamma}))^{3}+2n^{n}(det(P_{1}^{\Gamma}))(det(P_{2}^{\Gamma}))\nonumber\\ \leq && \frac{n^{\frac{3n-1}{2}}}{(n-1)^{\frac{n}{2}}}
\label{sepcond1}
\end{eqnarray}
Then the state $\varrho$ is positive partial transpose.\\
We will now give a counterexample to show that the converse of the above theorem is not true.
\subsection{Counterexample}
For this, let us consider a separable
 state described by the density operator $\varrho$
 \begin{eqnarray}\varrho=
 \begin{pmatrix}
  a_{1} & 0 & 0 & f_{1}^{*} \\
  0 & b_{1} & c_{1}^{*} & 0 \\
  0 & c_{1} & d_{1} & 0 \\
  f_{1} & 0 & 0 & e_{1}
  \end{pmatrix},
  \label{varrho}
\end{eqnarray}
where $a_{1}=\frac{1}{9}$, $b_{1}=\frac{1}{4}$, $c_{1}=\frac{1}{2\sqrt{117}}(1+i\sqrt{22})$, $d_{1}=\frac{23}{117}$,$e_{1}=\frac{23}{52}$, $f_{1}=\frac{1}{6\sqrt{13}}(1+i\sqrt{22})$. For the separable state $\varrho$, we find that $\varrho^{\Gamma}\neq P_{1}^{\Gamma}P_{2}^{\Gamma}$, where
$P_{1}^{\Gamma}P_{2}^{\Gamma}=
 \begin{pmatrix}
  a_{1} & 0 & 0 & f_{2}^{*} \\
  0 & b_{1} & c_{2}^{*} & 0 \\
  0 & c_{2} & e_{1} & 0 \\
  f_{2} & 0 & 0 & d_{1}
  \end{pmatrix}$
where $f_{2}=\frac{1}{3\sqrt{117}}(1+i\sqrt{22})$, $c_{2}=\frac{1}{4\sqrt{13}}(1+i\sqrt{22})$.\\
Therefore, the condition given in (\ref{decompositionrho2}) is a necessary condition for the separability
of two-qubit X-state.
\section{Illustration}
In this section, we have constructed the family of a X state which satisfy the inequality (\ref{sepcond1}) and thus it represent the family of separable states.\\
Let us consider a X state described by the density operator $\rho_{X}$ and it can be expressed as
{\begin{eqnarray}\rho_{X}=
 \begin{pmatrix}
  \rho_{11} & 0 & 0 & \rho_{14} \\
  0 & \rho_{22} & \rho_{23} & 0 \\
  0 & \rho_{32} & \rho_{33} & 0 \\
  \rho_{41} & 0 & 0 & \rho_{44}
  \end{pmatrix}, \sum_{i=1}^{4} \rho_{ii} = 1
  \label{rho}
\end{eqnarray}}
The quantum state $\rho_{X}$ given in (\ref{rho}) is positive semi-definite if $\rho_{22}\rho_{33}\geq |\rho_{23}|^{2}$ and $\rho_{11}\rho_{44}\geq |\rho_{14}|^{2}$. It is also known as two-qubit X-state. It is an
important class of state in the sense that any arbitrary two-qubit state can be transformed to it
by the parametric form of a unitary transformation \cite{mendonca}. Also it has been shown that
any two-qubit pure or mixed state can be transformed to two-qubit X-state with same entanglement by an
entanglement-preserving unitary transformation \cite{hedemann}. The concurrence of the quantum state $\rho_{X}$ is given
by \cite{hedemann,yu}
{\begin{eqnarray}
C(\rho)= 2 max\{0, |\rho_{32}|-\sqrt{\rho_{11}\rho_{44}}, |\rho_{41}|-\sqrt{\rho_{22}\rho_{33}}\} \label{concurrence}
 \end{eqnarray}}
The two-qubit X-state described by the density operator $\rho_{X}$ can be decomposed as
\begin{eqnarray}
\rho= P_{1}P_{2} \label{decompositionrho}
 \end{eqnarray}
where $P_{1}$ and $P_{2}$ denoting two positive definite matrices.\\
From (\ref{P1P2}) and (\ref{decompositionrho}), we can express the entries of $\rho$ in terms of $t_{1}$, $t_{2}$, $v_{1}$, $v_{1}^{*}$,$v_{2}$,$v_{2}^{*}$, which is given by
{\begin{eqnarray}
&&\rho_{11}=t_{1}, \rho_{14}=t_1v_1, \rho_{22}=t_{2}, \rho_{23}=t_2v_2, \nonumber\\&&
\rho_{32}=t_2v_2^*, \rho_{33}=t_2, \rho_{41}=t_1v_1^*, \rho_{44}=t_1 \nonumber\\&&
t_1 + t_2 = \frac{1}{2}, |v_1| \leq 1, |v_2| \leq 1
\end{eqnarray}}
The concurrence of $\rho_{X}$ can be re-expressed in terms of $t_{1}$, $t_{2}$, $|v_{1}|$ and $|v_{2}|$ as
{\begin{eqnarray}
C(\rho_{X})= 2 max\{0, t_2|v_2| - t_1, t_1|v_1| - t_2\} \label{concurrence1}
\end{eqnarray}}
We find that the concurrence of the state $\rho$ is zero when ${|v_1|}\leq\frac{t_{2}}{t_{1}}$ and ${|v_2|}\leq\frac{t_{1}}{t_{2}}$ hold. For the zero concurrence states, the inequality $|v_1||v_2| \leq 1$ is satisfied. It is indeed true since, $|v_1|\leq 1$ and $|v_2|\leq 1$.\\
In particular, if $t_{1}=t_{2}=\frac{1}{4}$ then concurrence $C(\rho)$ vanishes and the relation $(P_{1}P_{2})^{\Gamma}=P_{1}^{\Gamma}P_{2}^{\Gamma}$ holds. Thus, there exist a subclass of X-states which is described by the density operator $\rho_{X}^{(1)}$ is given by
{\begin{eqnarray}\rho_{X}^{(1)}=
 \begin{pmatrix}
  \frac{1}{4} & 0 & 0 & \frac{v_{1}}{4} \\
  0 & \frac{1}{4} & \frac{v_{2}}{4} & 0 \\
  0 & \frac{v_{2}^{*}}{4} & \frac{1}{4} & 0 \\
  \frac{v_{1}^{*}}{4} & 0 & 0 & \frac{1}{4}
  \end{pmatrix},
  \label{rho1}
\end{eqnarray}}
The state $\rho_{X}^{(1)}$ can be decomposed as
 \begin{eqnarray}
 \rho_{X}^{(1)}=P_{1}^{(1)}P_{2}^{(1)}
 \label{separablecond1}
 \end{eqnarray}
 where
 {
\begin{eqnarray} P_{1}^{(1)}=
  \begin{pmatrix}
  \frac{1}{4} & 0 & 0 & 0   \\
  0 & \frac{1}{4} & 0 & 0 \\
  0 & 0 & \frac{1}{4} & 0 \\
  0 & 0 & 0 & \frac{1}{4}
  \end{pmatrix},
  P_{2}^{(1)}=
  \begin{pmatrix}
  \frac{1}{4} & 0 & 0 & \frac{v_1}{4}\\
  0 & \frac{1}{4} & \frac{v_{2}}{4} & 0 \\
  0 & \frac{v_2^*}{4} & \frac{1}{4} & 0 \\
  \frac{v_1^*}{4} & 0 & 0 & \frac{1}{4}
  \end{pmatrix}
  \label{P1}
\end{eqnarray}}
It can be easily verified from the figure that the inequality (\ref{sepcond4}) is satisfied for (\ref{decompositionrho2}).
\begin{figure}[h]
	\centering
	\includegraphics[scale=0.3]{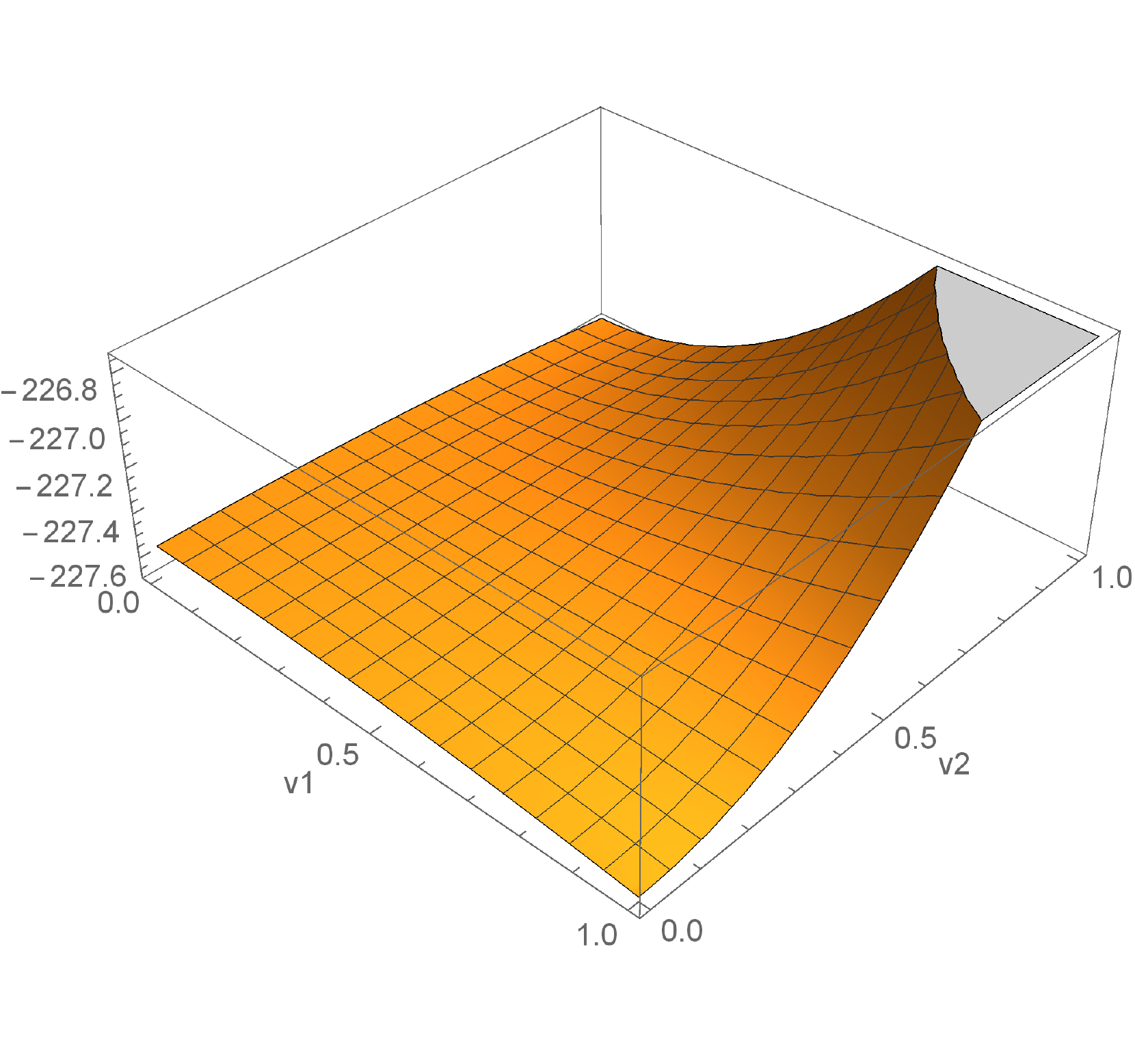}
	\caption{Verification of the inequality (\ref{sepcond4}) for $\rho_{X}^{(1)}=P_{1}^{(1)}P_{2}^{(1)}$}
\end{figure}

\section{Conclusion}
To summarize, we have explored the possibility of deriving the matrix equality $(AB)^{\Gamma}=A^{\Gamma}B^{\Gamma}$ for any $4 \times 4$ matrices $A$ and $B$, where $\Gamma$ denote the partial transposition. We found that, in general, the above matrix equality does not hold true for any $4 \times 4$ matrix. In particular, there exist special type of $4 \times 4$ matrices $A$ and $B$ for which $(AB)^{\Gamma}=A^{\Gamma}B^{\Gamma}$ is satisfied. We have studied and characterized those particular type of $4 \times 4$ matrices in the context of quantum information theory. As an application, we have shown that if there exist two positive semidefinite matrices $P_{1}$, $P_{2}$ such that $\sigma^{\Gamma}= (P_{1}P_{2})^{\Gamma}= P_{1}^{\Gamma}P_{2}^{\Gamma}$ holds and further (\ref{sepcond1}) is satisfied then a particular class of two-qubit X-states described by the density operator $\sigma$ is separable. Since our result can be applied to detect entanglement in two-qubit bipartite system so it gives the operational meaning of the derived matrix equality. Since the choice of the positive semidefinite matrices are not unique so the method developed to link the matrix equality $(AB)^{\Gamma}=A^{\Gamma}B^{\Gamma}$ and separability condition, can be generalised for any $4 \times 4$ density matrices. Due to the non-uniqueness property of the decomposition of the density matrix into two positive semi-definte matrices, there is a possibility to generalise the new procedure developed here and that may help in the detection of large classes of two-qubit and multi-qubit entangled states.

\end{document}